\documentclass[prx, notitlepage, twocolumn,nofootinbib,superscriptaddress,longbibliography]{revtex4-1}
\usepackage[colorlinks=true,citecolor=blue,linkcolor=blue]{hyperref}
\usepackage{graphicx}
\usepackage{color}
\usepackage{dcolumn}
\usepackage{bm}
\usepackage{mathrsfs}
\usepackage{amsmath}
\usepackage{amssymb}
\usepackage{amsthm}
\usepackage{amsfonts}
\usepackage{enumerate}
\usepackage{latexsym}
\usepackage{hyperref}
\usepackage{centernot}
\usepackage{multirow}

\begin{document}

\title{XFe$_{4}$Ge$_{2}$ (X = Y, Lu) and Mn$_{3}$Pt: Filling-enforced
magnetic topological metals}

\begin{abstract}
Magnetism, coupled with nontrivial band topology, can bring about many
interesting and exotic phenomena, so that magnetic topological materials
have attracted persistent research interest. However, compared with
non-magnetic topological materials (TMs), the magnetic TMs are less studied,
since their magnetic structures and topological phase transitions are
usually complex and the first-principles predictions are usually sensitive
on the effect of Coulomb interaction. In this work, we present a
comprehensive investigation of XFe$_{4}$Ge$_{2}$ (X = Y, Lu) and Mn$_{3}$Pt,
and find these materials to be filling-enforced magnetic topological metals.
Our first-principles calculations show that XFe$_{4}$Ge$_{2}$ (X = Y, Lu)
host Dirac points near the Fermi level at high symmetry point $S$. These
Dirac points are protected by $P\mathcal{T}$ symmetry ($P$ and $\mathcal{T}$
are inversion and time-reversal transformations, respectively) and a 2-fold screw
rotation symmetry. Moreover, through breaking $P\mathcal{T}$ symmetry, the
Dirac points would split into Weyl nodes. Mn$_{3}$Pt is found to host 4-fold degenerate band crossings in the whole high
symmetry path of $A$-$Z$. We also utilize the GGA + $U$ scheme to take into
account the effect of Coulomb repulsion and find that the filling-enforced
topological properties are naturally insensitive on $U$.
\end{abstract}

\date{\today }
\author{Di Wang}
\affiliation{National Laboratory of Solid State Microstructures and School of Physics,
Nanjing University, Nanjing 210093, China}
\affiliation{Collaborative Innovation Center of Advanced Microstructures, Nanjing
University, Nanjing 210093, China}
\author{Feng Tang}
\affiliation{National Laboratory of Solid State Microstructures and School of Physics,
Nanjing University, Nanjing 210093, China}
\affiliation{Collaborative Innovation Center of Advanced Microstructures, Nanjing
University, Nanjing 210093, China}
\author{Hoi Chun Po}
\affiliation{Department of Physics, Massachusetts Institute of Technology, Cambridge, MA,
USA}
\author{Ashvin Vishwanath}
\affiliation{Department of Physics, Harvard University, Cambridge, MA, USA}
\author{Xiangang Wan}
\affiliation{National Laboratory of Solid State Microstructures and School of Physics,
Nanjing University, Nanjing 210093, China}
\affiliation{Collaborative Innovation Center of Advanced Microstructures, Nanjing
University, Nanjing 210093, China}
\maketitle





\section{Introduction}

The topological nature of electronic bands has attracted tremendous
attention in condensed matter physics since the birth of topological
insulator \cite{ti-1,ti-2}. During the past decade, a variety of topological
materials have been discovered, including topological insulators \cite%
{ti-1,ti-2,B2C3-1,xia,B2C3-2}, Dirac semimetals \cite%
{dirac-1,dirac-2,dirac-3,dirac-4,dirac-5}, Weyl semimetals \cite%
{weyl-1,weyl-5,weyl-2}, node-line semimetals \cite%
{node-1,node-2,node-3,node-4,node-5}, topological crystalline insulators
\cite{TCI-0,TCI-1}, and various other topological phases \cite%
{hourglass,TCI-4,TCI-6,HOTI,highorder-2,Bansil,AshvinRev}. It is well-known
that the nontrivial band topology is usually protected by time-reversal ($%
\mathcal{T}$) symmetry or other spatial symmetries such as mirror symmetry,
glide symmetry, etc \cite{Bansil}. Furthermore, symmetry and band topology
are intertwined with each other, and as a result symmetry information can be used to
diagnose the band topology in a highly efficient manner. Nowadays the
time-reversal-invariant topological materials (TMs) have been extensively
studied in both theories and experiments. Recently, symmetry-based methods
for the efficient discovery of topological materials were developed \cite%
{Po-nc,tang-np,toponature-1-lilun,toponature-2-lilun}, and thousands of
nonmagnetic TM candidates have been proposed \cite%
{toponature-1,toponature-2,toponature-3}.

Compared to the time-reversal-invariant topological materials, magnetic
topological materials are also expected to show rich exotic phenomena \cite%
{mag-review}, such as\ axion insulators \cite%
{axionadd1,axionadd2,axion,EuIn2As2,MnBi2Te4-1,MnBi2Te4-a1,MnBi2Te4-2,Mn2Bi2Te5}%
, antiferromagnetic topological insulator \cite%
{mag-ti,MnBi2Te4-1,MnBi2Te4-a1,MnBi2Te4-2,Bi2MnSe4,Mn2Bi2Te5,EuSn2P2,Zhangyuanbo}%
, magnetic Dirac semimetal \cite{mag-dirac-1,mag-dirac-2,mag-dirac-3}, and
magnetic Weyl semimetals \cite{weyl-1,mag-weyl-2}. However, the predictions
on magnetic TMs are relatively rare and very few of them have been realized
in experiments up to now \cite{mag-review}. This limitation is originated
from the fact that the topological properties are usually accompanied by
significant spin-orbit coupling (SOC), while SOC typically leads to complex
magnetic structures which is difficult to characterize experimentally and
theoretically. Moreover, unlike non-magnetic systems, Coulomb interaction is
of substantial importance in most magnetic systems, and the Coulomb
repulsion is usually incorporated by the parameter $U$ in first principles
calculations. Therefore, the first-principles predictions for magnetic
topological materials usually depends on the value of $U$ \cite%
{axion,weyl-1,mag-weyl-2}.

Recently, the filling constraint for band insulator has been established to
discover topological semimetals \cite{filling2016,filling}. This method
enables the efficient search for filling-enforced topological materials
solely on their space group (SG) and the filling electron number. The
central logic is that there is a tight bound for fillings of band insulator
in a SG \cite{filling2016}. Once a material crystallizing in this SG own the
number of valence electrons per unit cell out of the tight bound, it cannot
be a insulator. Based on the filling-constraint method \cite%
{filling2016,filling}, one can readily calculate the filling constraint $\nu
_{BS}$ for a material according to its SG, where $\nu $ represents the filling
number of occupied electrons per unit cell. If $\nu $\ $\notin $ $\nu
_{BS}\cdot
\mathbb{Z}
$ (here $%
\mathbb{Z}
$ represents any integer), this material has an electron filling
incompatible with any band insulator, and it must have symmetry-protected
gaplessness near the Fermi energy (unless a further symmetry-breaking or correlated phase is realized). This type of material is referred to as
filling-enforced (semi)metals \cite{filling2016,filling}. Moreover, the
filling-constraint method has been extended to magnetic materials based on
their magnetic space group (MSG) symmetries \cite{1651}.\ Note that the
filling-constraint method is based on the interplay between electron filling
and (magnetic) space group symmetries, and therefore it is insensitive to
the precise value of $U$ so long as the relevant symmetries are preserved.

In this work, by applying the filling-constraint method \cite%
{filling2016,filling,1651}, we find serveral magnetic topological metals: XFe%
$_{4}$Ge$_{2}$ (X = Y, Lu) and Mn$_{3}$Pt. We display detailed analysis for
the topological features of XFe$_{4}$Ge$_{2}$, where Dirac points are
located at $S$ point near the Fermi energy. Moreover, the calculations show
that the Dirac points would split into Weyl nodes by a small perturbation.
We also perform the first-principles calculations for the high-temperature
phase of Mn$_{3}$Pt, where the bands along the $A$-$Z$ path are four fold
degenerate. The results show that the essential properties and our
conclusions do not depend on the value of $U$ as we expected.

\begin{figure}[tbp]
\centering\includegraphics[width=0.5\textwidth]{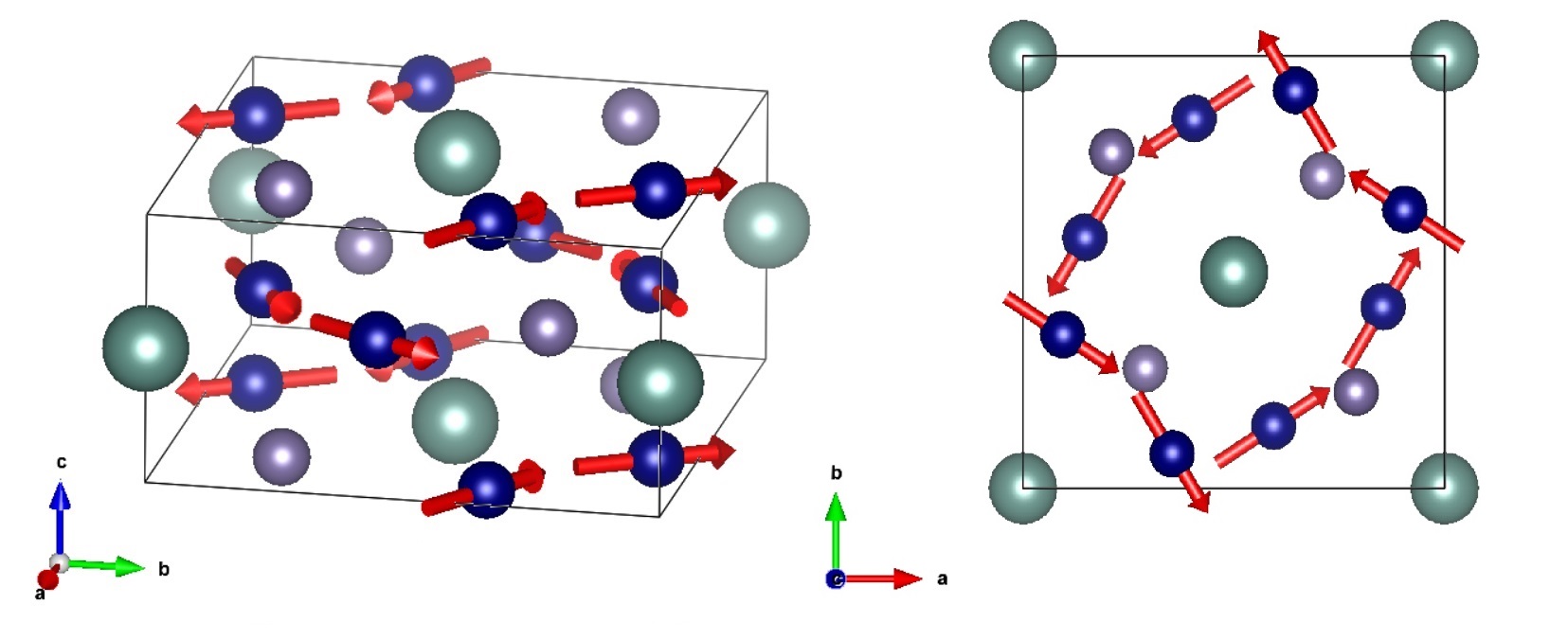}
\caption{Crystal structure of YFe$_{4}$Ge$_{2}$. The green, blue, and purple
balls represent the Y, Fe, and Ge ions, respectively. The arrows denote the
ground magnetic order measured by Ref. \protect\cite{exp-2001}.}
\label{crystal}
\end{figure}

\section{Method}

The calculations of electronic band structure and density of states have
been carried out as implemented in the Vienna ab-initio simulation package
(VASP) \cite{vasp1,vasp2,vasp3}. The Perdew--Burke--Ernzerhof (PBE) of
generalized gradient approximation (GGA) is chosen as the
exchange-correlation functional \cite{gga}. 6$\times $6$\times $12 and 16$%
\times $16$\times $8 k-point meshes are used for the Brillouin zone (BZ)
integral in XFe$_{4}$Ge$_{2}$ (X = Y, Lu) and Mn$_{3}$Pt system,
respectively. The self-consistent calculations are considered to be
converged when the difference in the total energy of the crystal does not
exceed 0.01 mRy. The effect of spin-orbit coupling (SOC) \cite{socref} is
considered self-consistently in all the calculations. We also utilize the
GGA + $U$ scheme \cite{LDA+U} to take into account the effect of Coulomb
repulsion in $3d$ orbital and the value of parameter $U$ is varied between 0
and 4 eV.

\section{Results}

\begin{figure}[tbp]
\centering\includegraphics[width=0.55\textwidth]{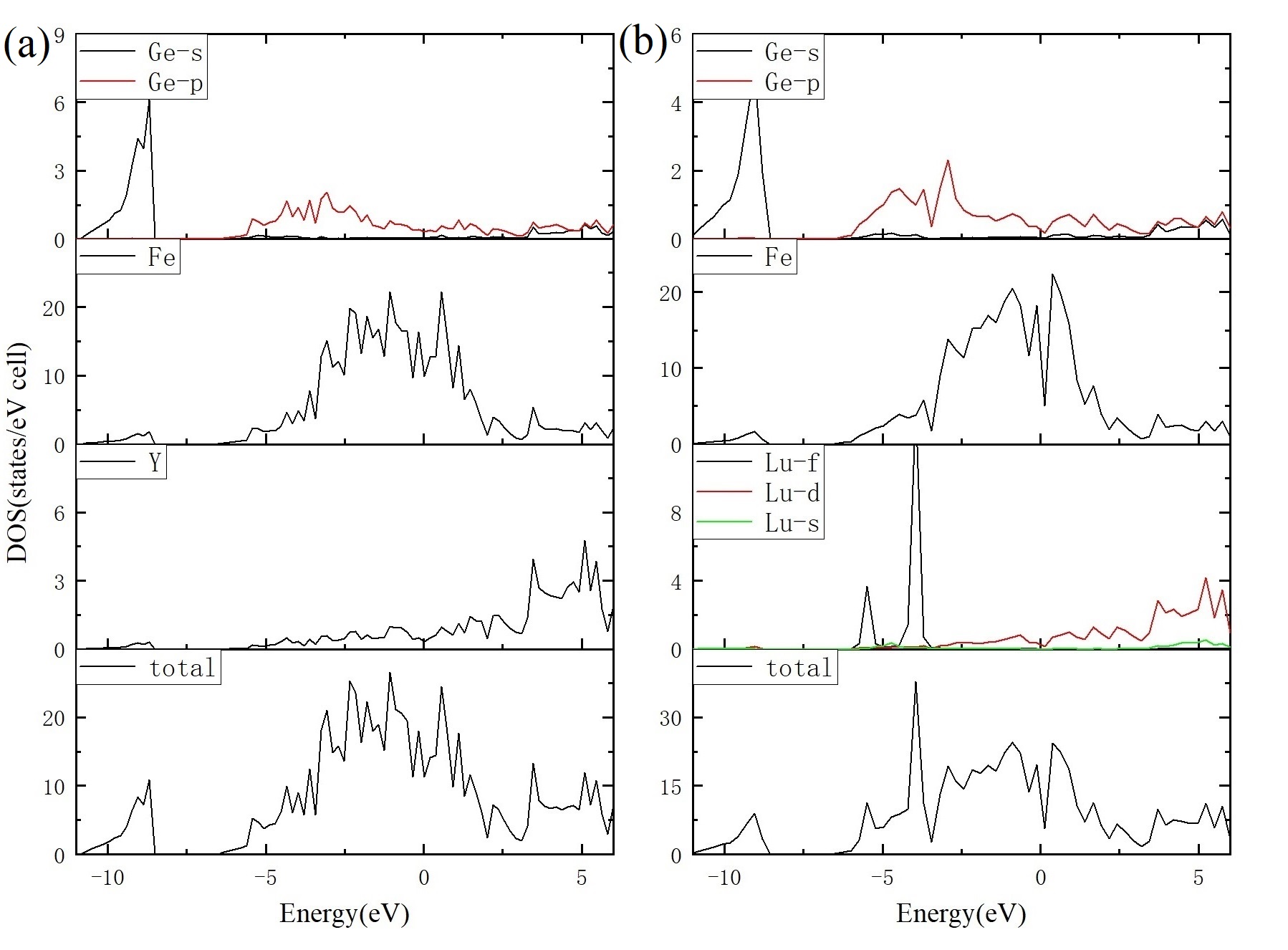}
\caption{Partial density of states (PDOS) of YFe$_{4}$Ge$_{2}$ (left) and
LuFe$_{4}$Ge$_{2}$ (right) from GGA calculation with non-collinear
antiferromagnetic configuration. The Fermi energy is set to zero. }
\label{dos1-1}
\end{figure}

\begin{figure*}[tbp]
\centering\includegraphics[width=0.9\textwidth]{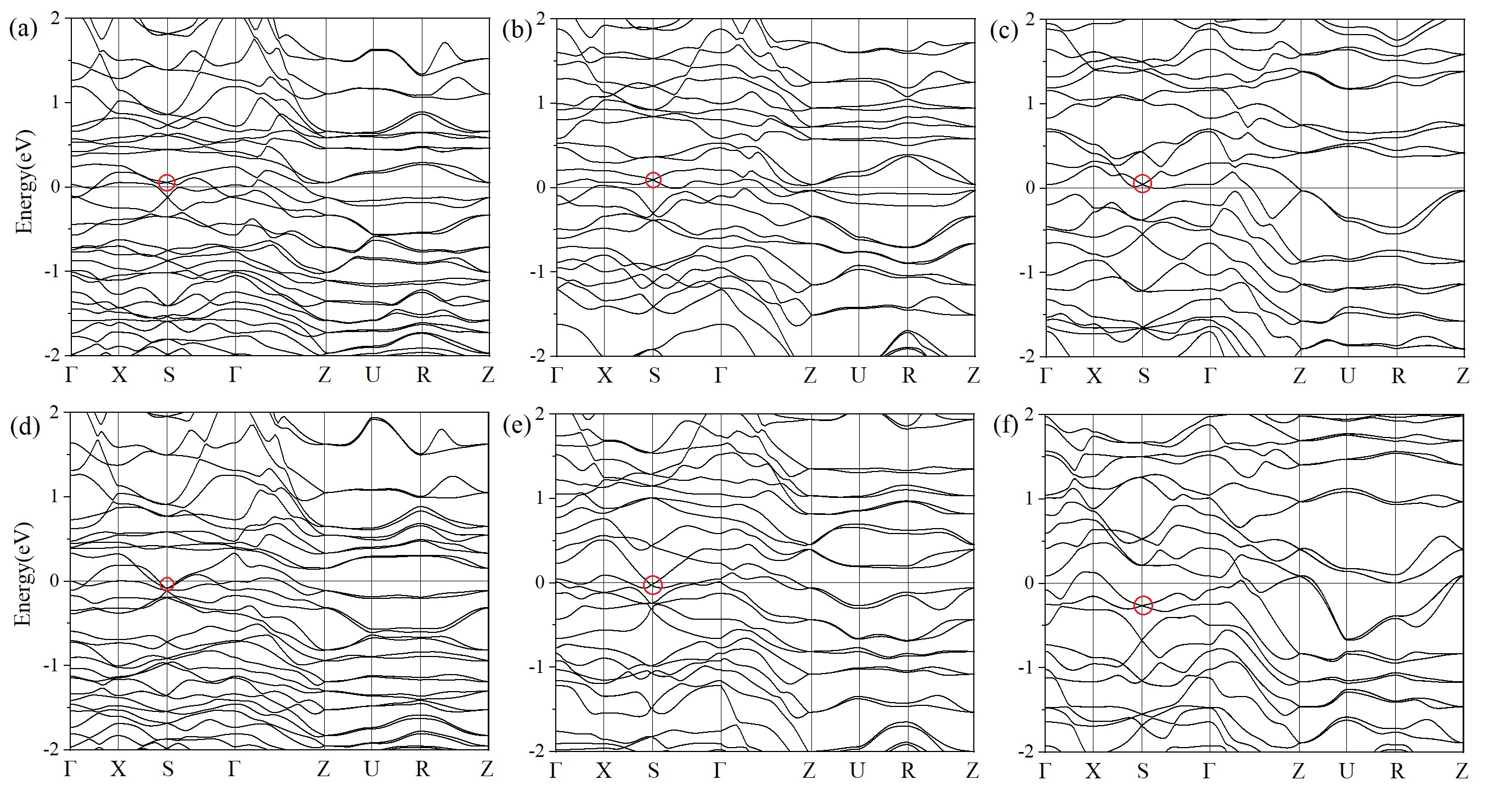}
\caption{Band structures of XFe$_{4}$Ge$_{2}$ (X = Y, Lu) with experimental
magnetic configuration \protect\cite{exp-2001}. (a)-(c) Band structures of
YFe$_{4}$Ge$_{2}$ from GGA, GGA+$U$ ($U$ = 2 eV) and GGA+$U$ ($U$ = 4 eV)
calculations, respectively. (d)-(f) Band structures of LuFe$_{4}$Ge$_{2}$
from GGA, GGA+$U$ ($U$ = 2 eV) and GGA+$U$ ($U$ = 4 eV) calculations,
respectively. The Fermi energy is set to zero. The Dirac points near the
Fermi level are marked with red circle.}
\label{bandstr}
\end{figure*}

YFe$_{4}$Ge$_{2}$ was previously reported to crystallize in the ZrFe$_{4}$Ge$%
_{2}$ type of structure with the space group $P4_{2}/mnm$\ at room
temperature \cite{exp-1975}. In this tetragonal structure, YFe$_{4}$Ge$_{2}$
has two formula units in the primitive unit cell \cite{exp-1975}. In 2001,
Schobinger-Papamantellos et al. \cite{exp-2001} measured the neutron
diffraction and magnetic properties of YFe$_{4}$Ge$_{2}$, and found a
magnetostructural (ferroelastic and antiferromagnetic) transition, where the
magnetic transition at T$_{N}$ = 43.5 K is accompanied by a first-order
phase transition from tetragonal structure ($P4_{2}/mnm$) to orthorhombic
structure ($Pnnm$). The magnetic structure below T$_{N}$\ is non-collinear\
antiferromagnetic with the type-III magnetic space group $Pn^{\prime
}n^{\prime }m^{\prime }$ (58.399 in the Belov-Neronova-Smirnova (BNS)
settings \cite{bns}), as shown in Fig. \ref{crystal}. Note that the magnetic
moments on two sites related by the inversion symmetry ($P$) point in
opposite directions, thus YFe$_{4}$Ge$_{2}$ is invariant under $P\mathcal{T}$
symmetry ($\mathcal{T}$ is the time-reversal transformation). The magnetic
moments of Fe ions at two sites are measured to be 0.63 $%
\mu
_{B}$ per Fe ion equally at 1.5 K. Similar to YFe$_{4}$Ge$_{2}$, LuFe$_{4}$Ge%
$_{2}$ has the first-order magnetoelastic transition at T$_{N}$ = 32 K from
non-magnetic tetragonal structure to antiferromagnetic orthorhombic
structure, while the Fe moment value is 0.45 $%
\mu
_{B}$ \cite{exp-2012}. These materials are suggestted to be filling-enforced
topological materials \cite{1651}, as we show in the following.

Based on the non-collinear antiferromagnetic configuration suggested by
neutron diffraction experiment \cite{exp-2001} as shown in Fig. \ref{crystal}%
, we perform the first-principles calculations for YFe$_{4}$Ge$_{2}$.
The density of states and the band structures are shown in Fig. \ref{dos1-1}
(a) and Fig. \ref{bandstr} (a)-(c), respectively. It should be noted that,
due to $P\mathcal{T}$ symmetry, the electronic bands in the whole BZ are
doubly degenerate, as shown in Fig. \ref{bandstr}. The bands in the energy
range from $-$10.0 and $-$8.0 eV are mainly contributed by Ge-$4s$ states,
while Y bands appear mainly above 3.0 eV. The $3d$ states of Fe ions are
mainly located from $-$6.0 to 2.0 eV, while Ge-$4p$ states appear mainly
between $-$6.0 to $-$2.0 eV, implying strong hybridization between Fe and Ge
states, as shown in Fig. \ref{dos1-1} (a). Due to hybridization between Fe
and Ge states, the Ge ion has a small calculated magnetic moment ($\sim $
0.05 $%
\mu
_{B}$), but the major magnetic moment is still located at the Fe site. Our
calculated magnetic moments of Fe ions incorporates all the symmetry
restrictions, and the absolute values of magnetic moments at two Fe sites
are 1.86 $%
\mu
_{B}$ and 1.70 $%
\mu
_{B}$, which is larger than the experimental value. Similar discrepancy has
also been reported in the calculations for other Fe-based intermetallic
compounds \cite{FeSe}. For YFe$_{4}$Ge$_{2}$ system, the filling constraint $%
\nu _{BS}$ for its MSG 58.399 is 4 \cite{filling2016,filling,1651}.
Meanwhile, the number of electrons per unit cell for YFe$_{4}$Ge$_{2}$ is to
$\nu $\ = 414, thus $\nu $\ $\notin $ $\nu _{BS}\cdot
\mathbb{Z}
$, indicating that YFe$_{4}$Ge$_{2}$ must be a filling-enforced material.

However, the filling-constraint method does not provide the detailed
topological properties. As shown in the symmetry analysis at $S$ point in
the next section, only a 4-dimensional irreducible representation is
allowed, thus all the states at the $S$ point must be grouped into Dirac
points. While this conclusion holds for all materials in the same magnetic
space group, the filling of YFe$_{4}$Ge$_{2}$ implies these Dirac points are
naturally close to the Fermi energy. As shown in Fig. \ref{bandstr} (a),
there is a Dirac point at only 56 meV above the Fermi level at $S$ point,
while the Dirac point below the Fermi level is relatively far away (at about
$-$120\ meV). We also take into account the effect of Coulomb repulsion in
Fe-$3d$ orbital by performing the GGA + $U$ calculations. The value of $U$
around 2 eV is commonly used in the Fe-based intermetallic compound \cite%
{U-01,U-02}. We have varied the value of $U$ from 0 to 4.0 eV ($U$ = 0 eV
represents GGA calculation without $U$), and the calculations show that the
position of the Dirac point is kept at the $S$ point with slightly varying
energy near the Fermi level, as shown in Fig. \ref{bandstr} (a)-(c). As
mentioned above, the filling-constraint method depends only on electron
filling and magnetic space group symmetries, thus the filling-enforced
topology is not sensitive to the calculation details.

We also perform the first-principles calculations of LuFe$_{4}$Ge$_{2}$
whose the band structures and the density of states are shown in Fig. \ref%
{bandstr} (d)-(f) and Fig. \ref{dos1-1} (b), respectively. Except Lu-$4f$
states which are located around $-$5 eV, the electronic properties of LuFe$%
_{4}$Ge$_{2}$ are very similar to YFe$_{4}$Ge$_{2}$, as shown in Fig. \ref%
{dos1-1} (b) and Fig. \ref{dos1-1} (a). The filling number of electrons per
unit cell is found to be $\nu $\ = 478,\ thus $\nu $\ $\notin $ $\nu
_{BS}\cdot
\mathbb{Z}
$, also identifying LuFe$_{4}$Ge$_{2}$ as featuring Dirac points pinned at $S$
point near the Fermi level.

\begin{figure}[tbp]
\centering\includegraphics[width=0.45\textwidth]{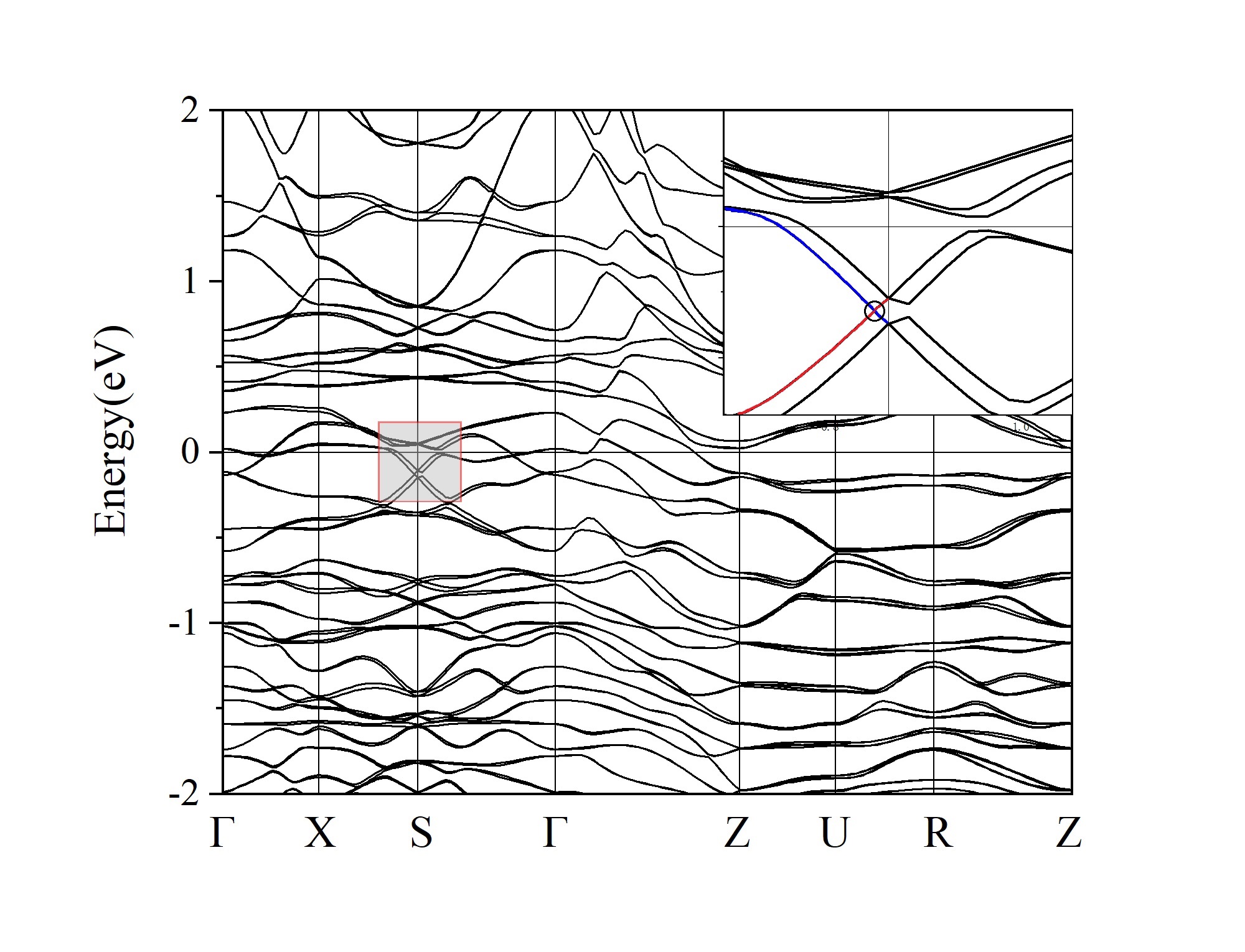}
\caption{Band structures of YFe$_{4}$Ge$_{2}$ from GGA calculation with the
magnetic configuration that is slightly deviated from the ground state. The
Fermi energy is set to zero. The inset is the detailed structure around $S$
point. The red and blue line represent the eigenstates of $%
\{2_{010}|1/2,1/2,1/2\}$ with the eigenvalues $-ie^{-i\protect\pi k_y}$ and $%
+ie^{-i\protect\pi k_y}$, respectively. The Weyl point along $X$-$S$ line are
marked with black circle.}
\label{picweyl}
\end{figure}

As mentioned above, YFe$_{4}$Ge$_{2}$ exhibits an antiferromagnetic order with
opposite spins related by inversion, and so $P\mathcal{T}$ symmetry is present and the electronic bands are doubly degenerate everywhere.
Upon breaking the $P\mathcal{T}$ symmetry, the Dirac cone may split into
Weyl nodes \cite{breakPT}. Note that, with the $P\mathcal{T}$ symmetry and the two-fold rotation $%
\{2_{001}|0\}$ symmetry coexisting in this system, the z-direction component
of Fe magnetic moment $m_{z}$ should be zero, and the magnetic moments are
lying in the xy-plane. By a small perturbation such as external field, the
magnetic configuration may have nonzero z-direction component with $P%
\mathcal{T}$ symmetry broken while $%
\{2_{001}|0\}$ preserved, and the Dirac cone may split into Weyl nodes.
Accordingly, we perform the GGA calculations with the magnetic state where
the magnetic moments have the deflection angle about 2$%
{{}^\circ}%
$ from xy-plane. As shown in Fig. \ref{picweyl}, at $S$ point, all the Dirac
points indeed split into Weyl points. In addition, there is also a symmetry-protected band crossing
in the path $X$-$S$, as shown in Fig. \ref{picweyl}. As discussed in the
symmetry analysis below, only a two-fold screw rotation $%
\{2_{010}|1/2,1/2,1/2\}$ is preserved for the $k$
point (1/2, $k_y$, 0) in the path $X$-$S$. The
first-principles results show that the red and blue line represent the
eigenstates of $\{2_{010}|1/2,1/2,1/2\}$ with the eigenvalues $-ie^{-i\pi k_y}$
and $+ie^{-i\pi k_y}$ respectively, as shown in Fig. \ref{picweyl}. Therefore
the hybridization between these two bands is forbidden and there is a
unavoidable crossing point located at $X$-$S$ line. Similarly, the
first-principles results show that there is also a unavoidable crossing
point along the path $Y$-$S$.

\begin{figure}[tbp]
\centering\includegraphics[width=0.45\textwidth]{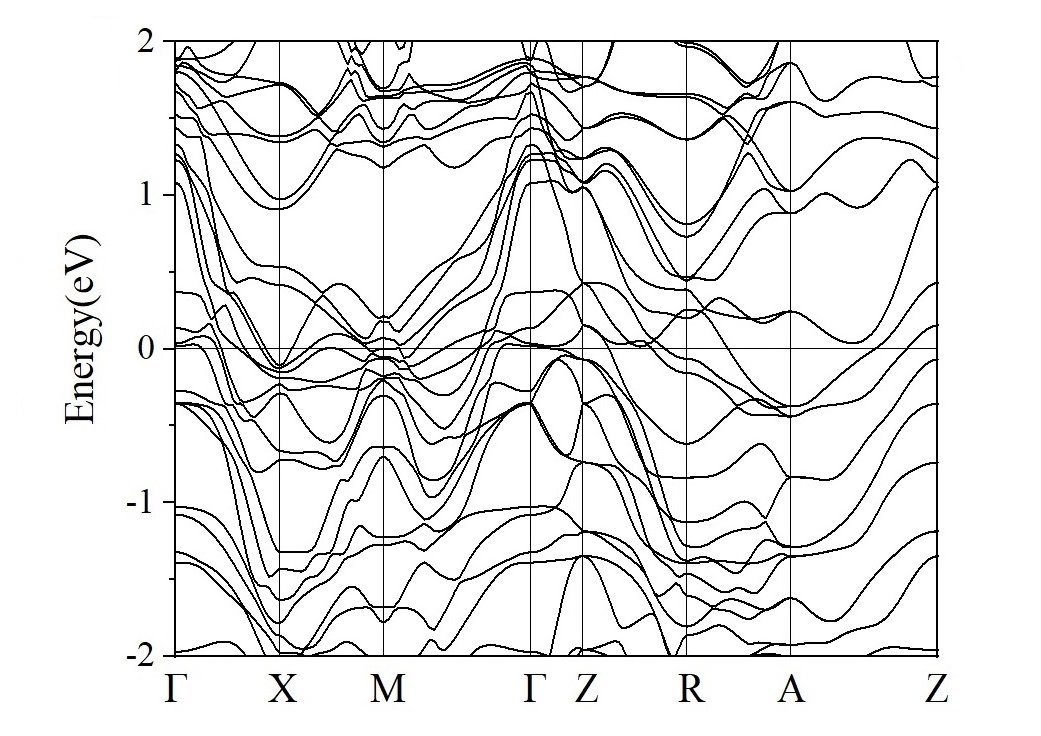}
\caption{Band structures of Mn$_{3}$Pt with high-temperature collinear
antiferromagnetic configuration from GGA calculation. The Fermi energy is
set to zero.}
\label{bandstr2}
\end{figure}

We also find the high-temperature phase of a cubic antiferromagnetic
intermetallic compound Mn$_{3}$Pt as another filling-enforced topological
material. Experiment reveals that Mn$_{3}$Pt crystallizes in a cubic crystal
structure (space group $Pm$-$3m$) at room temperature\ and has a long-range
antiferromagnetic order with T$_{N}$ = 475 K \cite%
{Mn3Pt-exp1,Mn3Pt-exp2,Mn3Pt-exp3}. Neutron diffraction experiments show a
first-order magnetic transition in Mn$_{3}$Pt system at about 365 K, between
a low-temperature non-collinear antiferromagnetic state and a
high-temperature collinear antiferromagnetic state \cite%
{Mn3Pt-exp1,Mn3Pt-exp2,Mn3Pt-exp3}. The high-temperature phase of Mn$_{3}$Pt
is collinear antiferromagnetic with the magnetic space group $P_{c}4_{2}/mcm$
(132.456), where the Mn atoms in xy-plane couple antiferromagnetically, and
the Mn atoms along z-direction also have opposite spin orientations. Very
recently, Liu et al. \cite{Mn3Pt-exp4} report the observation of the
anomalous Hall effect in thin films of the low-temperature phase for Mn$_{3}$%
Pt. They also show that the anomalous Hall effect can be turned on and off
by applying a small electric field at a temperature around 360 K and the Mn$%
_{3}$Pt is close to the phase transition \cite{Mn3Pt-exp4}. Therefore
exploring the possible exotic properties of the high-temperature phase for Mn%
$_{3}$Pt is also an interesting problem.

Similarly, we perform the first-principles calculations based on the
high-temperature phase of Mn$_{3}$Pt and the band structures are shown in
Fig. \ref{bandstr2}. The high-temperature phase of Mn$_{3}$Pt also has $P%
\mathcal{T}$ symmetry like YFe$_{4}$Ge$_{2}$, thus the electronic
band structures are symmetry-protected doubly degenerate in whole BZ. The
calculated magnetic moment at the Mn site is 2.9 $%
\mu
_{B}$ per Mn ion, which is in reasonable agreement with the experiment value
3.4 $%
\mu
_{B}$ \cite{Mn3Pt-exp1,Mn3Pt-exp2,Mn3Pt-exp3}. The Pt-$5d$ states are mainly
located from $-$6.0 to $-$3.0 eV. The $3d$ states of Mn ions appear
mainly from $-$3.0 to 2.0 eV. For Mn$_{3}$Pt, the filling number of electrons per
unit cell is 306 while the $\nu _{BS}$ for its MSG (132.456) is also 4, thus
$\nu $\ $\notin $ $\nu _{BS}\cdot
\mathbb{Z}
$, indicating that there is a half-occupied four-fold energy level near
Fermi energy. As shown in Fig. \ref{bandstr2}, the bands are four fold
degenerate in the path of $A$-$Z$, which is protected by the symmetry
operations of magnetic stucture, as shown in the detailed symmetry analysis
in the next section. We also vary the value of $U$ from 0 to 4.0 eV and find
that the four-fold energy level always exists.

\section{Symmetry analysis}

In this section, we show the detailed symmetry analysis for the Dirac band
crossings. We will first focus on the S point in XFe$_{4}$Ge$_{2}$ (X = Y, Lu), followed by a corresponding discussion for Mn$_3$Pt. For the $S$ point
(1/2, 1/2, 0), eight symmetry operations are generated by three symmetries:
the $P\mathcal{T}$ symmetry $\{-1^{\prime }|0\}$, a two-fold screw rotation $%
\{2_{100}|1/2,1/2,1/2\}$ and a two-fold rotation $\{2_{001}|0\}$, where the
left part represents the rotation and the right part means the lattice
translation. Note that $-1$ above denotes the inversion symmetry\ and the superscript prime
means an additional time-reversal operation $\mathcal{T}$ here. Since $%
\{2_{100}|1/2,1/2,1/2\}^{2}=-\{1|1,0,0\}$, where the minus sign originates
from the spin rotation,  the momentum phase factor equals to $-1$ for a
Bloch state at the $S$ point. Thus, the eigenvalues for $%
\{2_{100}|1/2,1/2,1/2\}$ is $\pm 1$. We can then choose the eigenstates $%
\psi _{nS}^{\pm }$ of $\{2_{100}|1/2,1/2,1/2\}$ at $S$ point, where the
superscript denotes the eigenvalue of $\{2_{100}|1/2,1/2,1/2\}$ and $n$ is
the band index. Because $[\{-1^{\prime }|0\},\{2_{100}|1/2,1/2,1/2\}]=0$ when acting on the Bloch states at S,
operation of $\{-1^{\prime }|0\}$ will preserve the eigenvalue of $%
\{2_{100}|1/2,1/2,1/2\}$ and result in the other state in the Kramers
doublet: i.e.\ $\{-1^{\prime }|0\}\psi _{nS}^{+}$ is orthogonal to $\psi
_{nS}^{+}$ but with the same eigenvalues of $\{2_{100}|1/2,1/2,1/2\}$.
Besides, it should be noted that $\{2_{100}|1/2,1/2,1/2\}\{2_{001}|0\}=-%
\{2_{001}|0\}\{2_{100}|1/2,1/2,1/2\}$, thus $\{2_{001}|0\}\psi _{nS}^{\pm }$
reverses the eigenvalue of $\{2_{100}|1/2,1/2,1/2\}$. So the four orthogonal states $\psi
_{nS}^{+},\{-1^{\prime }|0\}\psi _{nS}^{+},\{2_{001}|0\}\psi
_{nS}^{+},\{2_{001}|0\}\{-1^{\prime }|0\}\psi _{nS}^{+}$ are degenerate, constituting the basis of the
4-dimensional irreducible representation. Therefore, in XFe$_{4}$Ge$_{2}$
system, the $S$ point (1/2, 1/2, 0) only allow for 4-dimensional
irreducible representation.

Based on the $k\cdot p$ method, we build the
effective Hamiltonian by using the four relevant states as basis vectors, in
the order of $\psi _{nS}^{+},\{-1^{\prime }|0\}\psi
_{nS}^{+},\{2_{001}|0\}\psi _{nS}^{+},\{2_{001}|0\}\{-1^{\prime }|0\}\psi
_{nS}^{+}$. To the lowest order in $\mathbf{q}$, the Hamiltonian $\mathcal{H}%
(\mathbf{q})$ can be written as

\begin{equation}
\left[
\begin{array}{cccc}
\mathbf{q}_{x}C_{5} &  & \frac{-i\mathbf{q}_{z}C_{1}-\mathbf{q}_{y}C_{2}}{%
\sqrt{2}} & \frac{\mathbf{q}_{z}(iC_{3}-C_{4})}{\sqrt{2}} \\
& \mathbf{q}_{x}C_{5} & \frac{\mathbf{q}_{z}(iC_{3}+C_{4})}{\sqrt{2}} &
\frac{i\mathbf{q}_{z}C_{1}-\mathbf{q}_{y}C_{2}}{\sqrt{2}} \\
\frac{i\mathbf{q}_{z}C_{1}-\mathbf{q}_{y}C_{2}}{\sqrt{2}} & \frac{\mathbf{q}%
_{z}(-iC_{3}+C_{4})}{\sqrt{2}} & -\mathbf{q}_{x}C_{5} &  \\
\frac{\mathbf{q}_{z}(-iC_{3}-C_{4})}{\sqrt{2}} & \frac{-i\mathbf{q}_{z}C_{1}-%
\mathbf{q}_{y}C_{2}}{\sqrt{2}} &  & -\mathbf{q}_{x}C_{5}%
\end{array}%
\right] ,  \label{Hsw}
\end{equation}%
where $\mathbf{q}=\mathbf{k}-S$, and $C_{i}$ ($i$ = 1, 2, ..., 5) are
parameters. The effective Hamiltonian suggests a linear dispersion\ in
neighbourhood of $S$. It is worth mentioning that there is also only one
4-dimensional irreducible representation in $Z$ point (0, 0, 1/2), and the
dispersion around $Z$ point is also linear.

When $P\mathcal{T}$ symmetry is broken, there are only four symmetry
operations for the $S$ point (1/2, 1/2, 0) generated by $%
\{2_{100}|1/2,1/2,1/2\}$ and $\{2_{001}|0\}$. As mentioned before, for the
eigenstates $\psi _{nS}^{\pm }$ of $\{2_{100}|1/2,1/2,1/2\}$ at $S$ point, $%
\{2_{001}|0\}\psi _{nS}^{\pm }$ reverses the eigenvalue of $%
\{2_{100}|1/2,1/2,1/2\}$. So $\psi _{nS}^{+}\ $and $\{2_{001}|0\}\psi
_{nS}^{+}$ own the same energy and they are orthogonal with each other,
constituting the basis of the 2-dimensional irreducible representation.
Therefore all the states at the $S$ point must be grouped pairwise.
Meanwhile, for the $k$ point (1/2, $k_y$, 0) in the path $X$(1/2, 0, 0)$-S$(1/2,
1/2, 0), only a two-fold screw rotation $\{2_{010}|1/2,1/2,1/2\}$ is
preserved. Note that $\{2_{010}|1/2,1/2,1/2\}^{2}=-\{1|0,1,0\}$, for $k$
point (1/2, $k_y$, 0), the eigenvalues should be $\pm ie^{-i\pi k_y}$. As shown in
Fig. \ref{picweyl}, the first-principles results show that two bands along
this line belong to the eigenstates of $\{2_{010}|1/2,1/2,1/2\}$ with
different eigenvalues. Thus the hybridization between these two bands is
forbidden and the band crossing is symmetry protected, as shown in Fig. \ref%
{picweyl}. Similarly, for the $k$ point ($k_x$, 1/2, 0) in the path $Y$(0, 1/2,
0)$-S$(1/2, 1/2, 0), $\{2_{100}|1/2,1/2,1/2\}$ is preserved and the
first-principles results show that there is also a symmetry protected band
crossing along the $Y$-$S$\ path.

In Mn$_{3}$Pt, similar to the discussion above, we study the symmetry
operations of this magnetic structure and find that there is only one
4-dimensional irreducible representation along $A$-$Z$ path in the BZ: For
the path of $A(1/2,1/2,1/2)$-$Z(0,0,1/2)$, eight symmetry operations are
generated by three symmetries: the $P\mathcal{T}$ symmetry $\{-1^{\prime
}|0\}$, a two-fold screw rotation $\{2_{110}|0,0,1/2\}$ and a mirror
operation $\{m_{1-10}|0\}$. Since $\{2_{110}|0,0,1/2\}^{2}=-\{1|0,0,0\}$
(the minus sign is coming from the electron spin), the eigenvalues for $%
\{2_{110}|0,0,1/2\}$ is $\pm i$. We can then choose the eigenstates $\psi
_{nA-Z}^{\pm }$ of $\{2_{110}|0,0,1/2\}$ in $A$-$Z$ path, where the
superscript denotes the eigenvalue of $\pm i$ and $n$ is the band index.
Note that $\{-1^{\prime
}|0\}\{2_{110}|0,0,1/2\}=-\{2_{110}|0,0,1/2\}\{-1^{\prime }|0\}$, indicating
that $\{-1^{\prime }|0\}\psi _{nA-Z}^{+}$ is orthogonal to $\psi _{nA-Z}^{+}$
but with the same eigenvalues of $\{2_{110}|0,0,1/2\}.$ Besides, it should
be noted that $\{2_{110}|0,0,1/2\}\{m_{1-10}|0\}=-\{m_{1-10}|0\}%
\{2_{110}|0,0,1/2\}$, thus $\{m_{1-10}|0\}\psi _{nA-Z}^{\pm }$ reverses the
eigenvalue of $\{2_{110}|0,0,1/2\}$. So $\psi _{nA-Z}^{+}$, $\{-1^{\prime
}|0\}\psi _{nA-Z}^{+}$, $\{m_{1-10}|0\}\psi _{nA-Z}^{+}$, $%
\{m_{1-10}|0\}\{-1^{\prime }|0\}\psi _{nA-Z}^{+}$ are again orthogonal and degenerate, constituting the basis of the
4-dimensional irreducible representation. Therefore, in Mn$_{3}$Pt system,
only a 4-dimensional irreducible representation is allowed along the $%
A(1/2,1/2,1/2)$-$Z(0,0,1/2)$ path.

\section{Conclusion}

In conclusion, by applying the filling constraints, we discover several
magnetic topological semimetals: XFe$_{4}$Ge$_{2}$ (X = Y, Lu) and Mn$_{3}$%
Pt. The first-principles calculations show that YFe$_{4}$Ge$_{2}$ is a metal
with a Dirac cone located at $S$ point near the Fermi level, which is
protected by the symmetry operations of magnetic stucture. We have varied
the value of $U$ from 0 to 4.0 eV, and the results show that Dirac point
always exists, since the topological property is filling-enforced and
independent on $U$. When the magnetic moments have a small nonzero
z-direction component, the Dirac point would split into Weyl nodes around
 the $S$ point. We also perform the first-principles calculations
based on the high-temperature collinear antiferromagnetic configuration of Mn%
$_{3}$Pt. The calculation results and symmetry analysis show that it is also
a topological material.

\section{Acknowledgements}

DW, FT and XW were supported by the NSFC (No.11525417, 11834006, 51721001
and 11790311), National Key R\&D Program of China (No. 2018YFA0305704 and
2017YFA0303203) and the excellent programme in Nanjing University. AV was
supported by a Simons Investigator Award and by the Center for Advancement
of Topological Semimetals, an Energy Frontier Research Center funded by the
U.S. Department of Energy Office of Science, Office of Basic Energy
Sciences, through the Ames Laboratory under its Contract No.
DE-AC02-07CH11358. HCP was supported by a Pappalardo Fellowship at MIT and a
Croucher Foundation Fellowship.

\clearpage

\bibliography{XFe4Ge2_v15}

\end{document}